\documentstyle[aaspp4]{article}

\def\mincir{\raise -2.truept\hbox{\rlap{\hbox{$\sim$}}\raise5.truept
\hbox{$<$}\ }}

\def\magcir{\raise -2.truept\hbox{\rlap{\hbox{$\sim$}}\raise5.truept
\hbox{$>$}\ }}
\def\minmag{\raise-2.truept\hbox{\rlap{\hbox{$<$}}\raise 6.truept\hbox
{$>$}\ }}

  \def\Mpch{\, h^{-1}{\rm Mpc}}    

\def\Mix{C+2$\nu$DM}
\def\L{$\Lambda$CDM}
\def\L37{$\Lambda$CDM$_{0.3}$}    
\def\La{$\Lambda$CDM$_1$}
\def\Lb{$\Lambda$CDM$_2$}
\def\Lc{$\Lambda$CDM--80}
\def\dth{ \delta_{th} }

\begin{document}

\title{STATISTICAL TESTS FOR CHDM AND $\Lambda$CDM COSMOLOGIES }

\vglue 1.9truecm
\centerline{{\bf Sebastiano Ghigna}$^{1,2}$, {\bf Stefano Borgani}$^{3}$}

\centerline{{\bf Marco Tucci}$^{2}$, {\bf Silvio A. Bonometto}$^{2}$}

\centerline{ {\bf Anatoly Klypin }$^{4}$, {\bf Joel R. Primack}$^5$ }

\vglue 0.4truecm
\noindent
$^1$Department of Physics of Durham University, Science Laboratories, 
South Road, Durham DH1 3LE, England

\noindent
$^2$Dipartimento di Fisica dell'Universit\`a di Milano
Via Celoria 16, I-20133 Milano, Italy {\sl and }INFN, Sezione di Milano

\noindent
$^3$ INFN, Sezione di Perugia,
c/o Dipartimento di Fisica dell'Universit\`a,
Via A. Pascoli I-06100 Perugia, Italy

\noindent
$^4$Astronomy Department, New Mexico State University
Box 30001 -- Dept.~4500, Las Cruces, NM 88003-0001, USA
{\sl and }Astro-Space Center, Lebedev Phys. Inst., Moscow,
Russia

\noindent
$^5$Physics Department, University of California,
Santa Cruz, CA 95064, USA


\begin{abstract}

We apply several statistical estimators to high--resolution N--body
simulations of two currently viable cosmological models: a mixed dark
matter model, having $\Omega_\nu=0.2$ contributed by two massive neutrinos
(C$+2\nu$DM), and a Cold Dark Matter model with Cosmological Constant 
($\Lambda$CDM) with $\Omega_0=0.3$ and $h=0.7$. Our
aim is to compare simulated galaxy samples with the Perseus--Pisces
redshift survey (PPS). We consider the $n$--point correlation functions
($n=2$--$4$),  the $N$--count probability functions $P_N$, including the
void probability function $P_0$, and the underdensity probability
function $U_\epsilon$ (where $\epsilon$ fixes the underdensity threshold
in percentage of the average). We find that $P_0$ (for which PPS and CfA2 data
agree) and $P_1$ distinguish
efficiently between the models, while $U_\epsilon$ is only marginally
discriminatory. On the contrary, the reduced skewness 
and kurtosis are, respectively, $S_3\simeq 2.2$ and $S_4\simeq 6$--7 
in all cases, quite independent of the scale, in agreement with 
hierarchical scaling predictions and estimates based on redshift surveys.
Among our results, we emphasize the
remarkable agreement between PPS  data and C$+2\nu$DM in all the tests
performed. In contrast, 
the above $\Lambda$CDM model
has serious difficulties in reproducing observational data if galaxies and
matter overdensities are related in a simple way. 

\end{abstract}   


\section{Introduction}

Although many cosmological models have been considered by various
authors, much effort has been concentrated on models inspired by the
hypotheses of inflation. In such a context, 
the ``natural'' choice is to consider 
COBE normalized models with negligible spatial curvature, 
Gaussian and adiabatic
primordial fluctuations and a spectrum close to the Zel'dovich one.

Among them, models that agree with the available data
on large scales, where fluctuations are still in the linear
regime, deserve further inspection
at smaller scales. Critical differences
can be expected to exist for fluctuations still in a weakly non--linear regime.
Here we need statistical tests, which are 
simultaneously {\sl robust} and {\sl discriminatory},  to compare real
data with N--body simulations.  
On the
contrary, when much smaller scales are inspected, it is not clear whether
some residual signal coming from the shape of the post--recombination
spectrum can still be appreciated, and in any case it will be necessary to 
include astrophysics that is still poorly understood, such as star 
formation and feedback effects.

In a previous paper (Ghigna et al. 1994, hereafter paper G), we showed
that the void probability function (VPF), $P_0$, can be a robust and
discriminatory test on the distribution of matter in underdense regions
of the Universe,  which are however  in a weakly non--linear regime
($|\delta \rho| \mincir \bar \rho$). In paper G, voids in a volume--limited
sample of the Perseus--Pisces Survey (PPS, Giovanelli \& Haynes 1991)
were analysed and the function $P_0(r)$ was computed. 
This VPF was then
compared with simulations based on the $\Omega_0=1$ CDM model and on the 
CHDM model with $\Omega_\nu=0.3$ and one massive neutrino
($m_\nu \simeq 7\, $eV). We found that this CHDM model produces too 
many intermediate-size voids.

In this paper we extend the comparison between PPS data and simulations
by considering new models and new tests. We study a $\Lambda$CDM model
with $\Omega_0 =0.3$ and $h=0.7$ 
(hereafter this model will be
called \L37), and a mixed dark matter  model with
$\Omega_\nu=0.2$, $h=0.5$, and 2  massive neutrinos, each having a mass of 
$m_\nu = 2.3\, $eV (hereafter this mix will be called C+2$\nu$DM).  
The \L37 model has been found to reproduce the power--spectrum shape on 
intermediate ($\sim 20\,h^{-1}$Mpc) scales (e.g., Peacock \& 
Dodds 1994; Borgani et al. 1996), as well as the abundance of galaxy clusters
(e.g., Eke et al. 1996, and references therein). Klypin, Primack, \& Holtzman 
(1996) showed however that it predicts a too strong galaxy 
clustering on scales $\mincir 10\,h^{-1}$Mpc. As for the C+2$\nu$DM model, 
it was 
firstly considered by Primack et al. (1995). Sharing $\Omega_\nu$ between 
two massive $\nu$ species decreases the fluctuation amplitude on $\simeq
10 \,h^{-1}$Mpc scales with respect to the standard CHDM model (thus
alleviating cluster overproduction), without reducing the
small--scale ($\sim 1\,h^{-1}$Mpc) power to an unacceptable level for
early galaxy formation.

Among previous works on the VPF, it should 
be mentioned that Fry et al. (1989) estimated it for a
preliminary version of PPS and compared the results with CDM N--body 
simulations. Weinberg \& Cole (1992) showed that the VPF
can also discriminate between Gaussian and non--Gaussian initial
conditions. Finally, Ghigna et al. (1996) compared the void statistics 
in the PPS
sample analyzed here and to simulations of the Broken Scale Invariance (BSI) 
model (CDM with a characteristic scale in the post--inflationary spectrum; 
see, e.g., Gottl\"ober, M\"ucket \& Starobinski 1994), 
which was found to agree with observations. 

The $n$--point correlation functions $\bar \xi_n$
for the PPS, evaluated through the 
counts--of--neighbours technique,  were also considered 
in previous papers (Bonometto et al. 1993 and 1995; Ghigna et al. 1996),
but their comparison with
N--body simulations did not show a strong discriminatory power. Here we
work them out through a different technique (counts--in--cells), which
however confirms previous results.

In addition to the $n$--point correlation functions and the VPF, we address  
the $N$--count probability functions $P_N (r)$ and the underdensity function
$U_\epsilon (r)$, defined as the probability 
that a randomly placed sphere has a
galaxy density below $\epsilon$\% of the average.  The functions $P_N
(r)$ ($N\ge1$) and $U_\epsilon (r)$ were never estimated before in
samples with depth comparable with PPS.

PPS results are given here in the CMB reference frame (at variance with
paper G, where the volume--limited subsample was worked out in the local group
rest frame), and are therefore more suitable for comparison with our
simulations. As a matter of fact, however, there is hardly any difference
between the analyses realized in the two frames (filled circles for LG vs.
triangles for CMB in Figure~1). 

Void analyses were also performed on the CfA2 survey (Vogeley, Geller, \&
Huchra 1991; Vogeley et al. 1994), the SSRS (Maurogordato, Schaeffer, \&
da Costa 1992; open circles in Figure 1) 
and the 1.2 Jy IRAS redshift survey (Bouchet et al. 1993).
It is remarkable that these results are globally consistent with the 
PPS ones, despite the fact that they refer to samples defined in 
different ways and differently located in the sky.

A crucial point, when testing cosmological models through 
the VPF statistics concerns the
identification of  galaxies. Indeed, a change in the efficiency of galaxy
formation in underdense regions has an immediate impact on  $P_0$
(Betancort--Rijo 1990; Einasto et al. 1991; Little \& Weinberg 1994). The
last authors explored three different criteria to identify galaxies in
$N$--body simulations: (i) as peak--particles of the linear density field
(e.g., Davis et al. 1985), (ii) using the  biasing relation derived by
Cen \& Ostriker (1993) from their CDM hydrodynamic simulations, (iii) as
high--density regions in the evolved density field. 
However,  some concerns have been raised about 
whether the linear biasing approach
yields the seeds where non--linear structures later form (e.g., Katz,
Quinn, \& Gelb 1993). Furthermore, although physically motivated, the Cen
\& Ostriker results were from CDM simulations performed within a
limited dynamical range.  For these reasons, we decided to identify
galaxies as corresponding
to high peaks of the evolved density field. However, even within this
choice, different criteria to fragment overmerged structures into
individual objects can be proposed (e.g., Gelb \& Bertschinger 1993). As
a general criterion, galaxy identification should be required to produce
the basic observed properties of the galaxy distribution, i.e. their
average separation, two--point correlation function and, possibly,
the observed luminosity function. In the next Section,  we will discuss the
simple technique used here to identify galaxies and compare it with the
approach adopted in paper G. 

Based on the simulation outputs, we generate 
artificial samples in redshift space
having the same geometry and number of galaxies of the
volume--limited sample extracted from PPS. 
We extract several samples from each simulation box
corresponding to different viewpoints, so as to obtain an estimate of the 
{\sl sky variance} within a given real--space volume.

\section{Real and simulated data sets.}

\vglue 0.2truecm
\noindent
{\sl Real data. --} The PPS database (Giovanelli \& Haynes 1989 and 1991)
is  limited to the region bound by $22^h \le \alpha \le 3^h\; 10^m$,
$0^\circ \le \delta \le 42^\circ\; 30^\prime$ to avoid areas of high
galactic extinction. Zwicky magnitudes of all galaxies brighter than
$m_{Zw}=15.5$ are however corrected for extinction, by using the
absorption maps of Burstein \& Heiles (1978). The resulting sample
includes 3395 galaxies and is virtually 100\% complete for all
morphological types up to $m_{Zw}=15.5$.  Observed velocities are then
corrected by subtracting the component of our 
velocity relative to the CMB, therefore  putting the
observer at rest in the CMB frame. A volume--limited subsample (VLS) is
then extracted, whose limiting magnitude $M_{\rm lim} = -19+5\log h$
corresponds to a limiting depth of $79\, h^{-1}$Mpc. This sample contains
902 galaxies with mean galaxy separation $d=5.5\,h^{-1}$ Mpc. This
sample differs from the one used in Paper~G, which was obtained by setting
the observer at rest with respect to the centroid of the Local Group. The
presence of a large volume, moving coherently with the local group,
allowed us to include several middle distance faint galaxies in the old
volume--limited sample. Hence the total number of galaxies it
contained was 1032, and their average separation was
$5.2\,h^{-1}$ Mpc. The decrease of the number of galaxies and
consequent slightly worse statistics, is the price to be paid to have
full coherence between observed and simulated data. 

\vskip 0.3truecm
\noindent
{\sl Simulated samples. --} We used four different PM simulations
obtained evolving $256^3$ cold particles on a 800$^3$ cell grid. More in
detail: (i) One realization of C+2$\nu$DM, with an additional $2\times 256^3$ 
hot particles, in a box with side
$l=50\,h^{-1}$Mpc ($h  = 0.5$), normalized to a COBE quadrupole $Q_{rms-PS} =
17\, \mu$K and yielding $\sigma_8 = 0.67$ 
(Primack et al. 1995). (ii)--(iii) Two realizations of
\L37\ (\La\ and \Lb)
 in a box with side $l=50 h^{-1}$Mpc ($h = 0.7$), normalized
to a COBE quadrupole $Q_{rms-PS} = 21.6\, \mu$K and yielding $\sigma_8 = 1.10$.
The first one started from the same random numbers as C+2$\nu$DM. (iv) A
further realization of \L37 in a box with side $l=80 h^{-1}$Mpc (\Lc),
with the same  normalization as above. (The \L37 simulations are from
Klypin, Primack \& Holtzman 1996). All these models assume a primordial 
spectral index $n=1$.

Let us now discuss the criteria followed to identify galaxies. As in paper G,
galaxies are set in overdensities exceeding a given threshold, but here 
the simulation output was preliminarily treated in such a way as 
to provide a direct individuation of overdensity regions. In paper G,
we had first found the number $n_p$ of particles in each cell
to single out local density maxima. However, single cells
are below the resolution allowed by PM codes. So now, the density of each
cell has been gauged by considering the sum $\sum_{p=1}^{27} n_p$ of the 
particles contained in a $3 \times 3 \times 3$ cell box centered on each cell.
Here, the simulation output, in addition to listing coordinates
and velocities  for DM particles, also gives us directly the density contrast
$\delta$ in a 27--cell volume centered on each of them (actually, 
we use a large random subsample of particles with uniform probability, 
amounting to a 20\% fraction of the total).

Therefore, we can simply select {\sl a priori} a threshold density contrast
$\delta_{th}$ and consider only particles with $\delta \ge \delta_{\rm th}$.
We considered three values: $\delta_{th}=$100, 150, 400. 
They are large enough to ensure that peaks above threshold
correspond to virialized structures. 
The two lowest values are (more and less) conservative estimates of 
the typical density 
constrast associated with structures becoming virialized at the present
epoch, while 
the highest value allows us to significantly perturb this basic
distribution of objects. 

The total numbers
of particles above the $\delta_{\rm th}$ selected are still quite large,
as expected (about 5--$10\, \%$ of the total).
Among them we select a small subset at random (751 particles 
for the box of side $l = 50\, h^{-1}$Mpc and 3077 for the box of side
$l = 80\, h^{-1}$Mpc), in order that the average inter--particle separation is
$d_{\rm gal}=5.5\Mpch$, i.e. the average galaxy separation 
in the real volume--limited sample. 

By construction, the {\sl surviving} particles are located in
 regions whose overdensities are above the thresholds selected  and
the distributions of those particles inside the {\sl parent} 
overdensities automatically fit
the different density profiles of such regions within the ``noise
interval''
introduced by the randomization process, which anyway should be expected
to occur in the real world as well.

In principle, passing from $\sim 1/10$ of DM particles down to $\sim 1/1000$
could introduce a bias, namely when small overdensity regions are considered.
The volume--limited sample extracted from PPS contains galaxies with
luminosity exceeding $L_* \simeq 10^{10}h^{-2}L_\odot$. Accordingly,
overdensity regions whose mass is $\sim 10^{12}h^{-1} M_\odot$, and typically
yield one or two galaxies, can be {\sl casually} included or excluded from the
artificial samples. This point is potentially delicate, especially for measures
like VPF, whose output could be affected by the inclusion or exclusion of a few
isolated galaxies. 

We addressed much care to this point, by building artificial samples from 
different random choices and comparing the outputs of our statistical
measures for them, although most results reported in this paper, for the sake 
of homogeneity, come from a single realization. In the next Section
we will debate this point further. We only anticipate here 
that the effect of changing the random subset of particles
is always quite modest, apart from a few cases whose anomaly is apparent.
Moreover, the scatter induced by such an effect is smaller than that
associated to the change of the observer setting within a given realization.
The results reported
were however checked to be {\sl typical}, by comparing  10 different
realizations. 

As a further check of the robustness of the results based on the above
galaxy identification method, we implemented in the C+2$\nu$DM simulation
two further prescriptions, both starting from the identification of {\em 
DM halos} and, therefore, free of this possible source of bias.
Such prescriptions were also meant to approach the procedure followed in
paper G, in spite of the different characteristics of the simulations
used here. The method of this countercheck is described in the
next paragraph and the results are reported in the next Section (cf.
Figure~7);
they confirm the validity of our standard procedure.

As a starting point to identify halos, we select
local density maxima on the grid, whose overdensity is greater than 200.
Afterwards, we center a sphere on this point, with radius equal to that at
which the overdensity drops to 200. The center of mass of the cold particles
falling within the sphere is then computed and used as the starting point for
the next iteration. We always find that this procedure converges after few
iterations. At the end, the mass of the 
halo is defined as the sum of the masses of all the member DM particles.
The resulting sample of DM halos is then used to identify galaxies.
The two prescriptions correspond then to two extreme cases:
\begin{description}
\item[(i)] No fragmentation: $N_{gal}=(l/d_{gal})^3=750$ galaxies are
identified as the $N_{gal}$ most massive halos. Each halo is then
identified with a single galaxy. The resulting halo mass threshold is
$M_{th}\simeq1.5\times 10^{12}h^{-1}M_\odot$.
\item[(ii)] Fragmentation: in order to break up halos we follow the same
simple prescription described by Bonometto et al. (1995). After a halo
mass threshold is chosen, the number $N_i$ of galaxies belonging to the 
$i$-th halo of mass $M_i$ is assumed to be $N_i=[M_i/M_{th}]$, where
$[x]$ denotes the largest integer that does not exceed $x$.
Therefore, the resulting mass threshold, $M_{th}\simeq 2.4\times
10^{12}h^{-1}M_\odot$,  is fixed by requiring that the total number
of galaxies matches $N_{gal}$. Fragments are assigned random positions
within the radius of the parent halo and velocities drawn from a Gaussian
distribution having mean equal to the halo peculiar velocity and
dispersion equal to the rms velocity of the member cold particles.
\end{description} 

As already outlined and discussed in paper G, these two prescriptions represent
extreme cases within a class of fragmentation methods not relying on
local anti--biasing. Therefore, although we do not attach to them any strong
physical motivation, they can be reliably used for bracketing results based on
more refined approaches. 

At variance with paper G, both the standard procedure used in the present
work, and the two latter prescriptions, make no recourse to the galaxy
luminosity function. In the simplest way, this would require one
additional parameter, the mass--to--light ratio $M/L$ of overdensity
regions,  which cannot be easily related to the physical $M/L$ of
well--defined objects and generally would depend on the resolutions of
the simulations (see also Ghigna et al. 1996). The outputs are however
strictly analogous.  In conclusion, what we work out are {\sl galaxies},
located in overdensity regions, with suitable individual velocities which
are essential to set them in redshift space. Overdensities were verified
to be essentially in virial equilibrium. Henceforth, the velocity
distribution for each region above threshold is quite similar to the one
considered in paper G, where each cell above threshold was given a total
galaxy mass proportional to $\sum_{p=1}^{27} n_p$ and  virial equilibrium
was explicitly imposed to obtain individual galaxy velocities.  
As a final consideration, let us notice that these procedures, as well as
the one adopted in paper G, do not leave room for any form of velocity
bias.  It is however important to notice that, also thanks to that, we
were able to keep the number of parameters fixing the distribution down
to one.

\vskip 0.3truecm
\noindent
{\sl Data--simulation comparison. --} The comparison between real and simulated
data is performed in redshift space, by extracting from the periodic 
simulation box (with replication)
a volume with the same geometry and measures of the real PPS, with respect to a
given {\sl observer} setting. 
More details about this operation are given in paper G (see also Ghigna
et al. 1996). The main question is related
to the fact that the volume--limited sample has a depth of $79\, h^{-1}$Mpc,
while the smaller simulation boxes have a side $l = 50\, h^{-1}$Mpc. As already
outlined in paper~G, this is not a real problem, since our analysis concerns 
scales $\mincir 13\Mpch$.

For each simulation and each choice of $\delta_{\rm th}$ 
we considered several observer settings, by selecting at random both the
location of the observer and the direction of the axis of the volume
observed. However, for each random setting, we first verified that the 
galaxy density 
in the artificial PPS sample differed by less than 2\% from 
the expected one ($=902/V_{VLS}$).  
 In this way 5 different observer setting were selected for each case.
As we shall see below, the scatter among observers, which is a measure of 
the {\sl sky--variance}, is always small and approximately 
of the same order as bootstrap errors. 

\section{Statistical analyses}

We estimate the statistical distribution of galaxies in each sample
through the count--in--cell technique. We work out the
probabilities $P_N$ that a randomly placed
cell contains $N$ galaxies. From this we
compute the moments of counts and obtain the volume--averaged 
correlation functions $\bar\xi_n$, after subtracting shot--noise
contributions (see Bonometto et al. 1995, for more details).
As in paper~G we use spherical cells completely contained in the sample
boundaries whose radii
$R$ are in the range 1--13$\Mpch$ and, at each $R$, 
we take $N_R=2\,V_{VLS}/V_R$ spheres randomly distributed in the
sample volume. Here $V_{VLS}\simeq 1.5\times 10^5 h^{-3}$Mpc$^3$ is 
the volume 
of the sample and $V_R=4\pi R^3/3$. As suggested by Fry \& Gazta\~naga
(1994), $N_R$ should give a sensible 
estimate of the number of independent cells
that can be allocated in the volume $V_{VLS}$ in the presence of clustering
(therefore the factor 2). This argument works well 
at least at relatively large scales for which $\bar\xi_2(R) \mincir 1$. At 
smaller $R$, underestimating
$N_R$ can in principle make the outcome of a measure excessively 
dependent on the set of $N_R$ spheres chosen. We verified 
that this is not the case, by analyzing the PPS sample for 20 different 
realizations of the positions of the spheres. The small shifts occuring
at the smallest radii are anyway accounted for by our estimates of
errors, which we obtain through the bootstrap resampling technique
(e.g. Ling, Frenk \& Barrow 1986). We consider up to 50 resamplings,
even though we find rapid convergence and a value of 20 would already provide
satisfactory estimates. In the following figures, for reasons
of clarity, bootstrap errors will be reported only for the observational 
data, but they also affect the results on simulated data, with similar 
magnitudes. (For a careful analysis of the uncertainties 
in count--in--cell statistics 
see Colombi, Bouchet \& Schaffer 1994, 1995 and Szapudi \& Colombi 1996.)

For each galaxy  sample, 
we worked out $\bar\xi_n(R)$ for $n=2$,3,4, i.e. variance, skewness
and kurtosis respectively. As far as $\bar\xi_{3,4}$ are concerned,
we will refer to the {\sl reduced} cumulants $S_3
\equiv \bar\xi_3/\bar\xi_2^2$  and $S_4 \equiv\bar\xi_4/\bar\xi_2^3$.

As for the $P_N$s, we examined them up to $N=5$ and $U_\epsilon$ for
$\epsilon$ in the range 30--70$\%$, but for $N\ge2$ and $\epsilon> 30\%$ the
discriminatory power is virtually absent. Values of $\epsilon$ less than
$30$ are hardly distinguishable from $P_0$ over most of the range of scales
considered. For these reasons, we will report results only for $P_0$, 
$P_1$ and $U_{30}$. 

As mentioned in the Introduction, an important point concerning the general 
significance of our analysis is whether
the PPS catalogue provides us with a fair sample of the Universe. Although we 
cannot give an answer to this question, we can at least check its reliability
against similar data available in the literature for other galaxy surveys.

In Figure~1 we compare the results of our PPS analysis on 
$\bar \xi_2(R)$
and VPF both in the CMB (triangles) and in the LG (filled circles) frame 
with that by Vogeley et al. (1994) for a volume--limited subsample of the 
CfA2 survey, having the same limiting magnitude of the PPS VLS that we consider
(open circles; the average between northern and southern CfA2
samples is plotted here). In order to be consistent with the analysis by
Vogeley et al., their results must be compared to those of PPS in the LG frame.
Therefore,
only for the purpose of this comparison, we resort to the same version of
the PPS sample as considered in paper G. 
In any case, it turns out that results are very weakly dependent on the 
frame in which redshifts are measured. Errorbars for PPS correspond to
$3\sigma$ bootstrap uncertainties. It is remarkable how close
the results for the two surveys are over the whole explored scale range, thus
indicating that PPS and CfA2 are essentially equivalent for the
statistical analyses we are considering here.

In Figure 2 we plot $\bar \xi_2(R)$ for the four simulations considered
at the three overdensity thresholds, $\delta_{th}$$=100$, 150, 
400 (note that the finite volume of the simulation affects scales
$\magcir 10 \Mpch$). The curves are obtained by averaging over 
5 observer locations. The
points (filled circles) refer to the PPS sample and their 
errorbars are $3\, \sigma$ from 20 bootstrap resamplings of PPS data. The
C+2$\nu$DM model clearly provides a good fit to the observational data,
especially for $\delta_{th}$$=150$
and an even better fit could be obtained by 
setting $\delta_{th}$$=180$. Let us recall that this is roughly
the density contrast expected for a virialized system in the 
approximation of spherical collapse.
On the contrary, the artificial galaxy samples that we extract from the 
\L37 simulations do not reproduce PPS data. Both the
amplitude and the slope of $\bar \xi_2$ are not satisfactory. Moreover, 
the dependence on
$\delta_{th}$ seems weaker here than for C+2$\nu$DM.
The effect of sky--variance 
can be seen in Figure~3, which shows 
again  how hard it is to find an observer setting in \L37 whose sky has
a $\bar\xi_2$ consistent with the PPS one. These difficulties 
of \L37 were however already known (Klypin, Primack \&
Holtzman 1996).

The dependence of $S_3$ on $R$ is shown in Figure~4, where, as usual,
errorbars are $3\sigma$ for PPS data
and curves refer to simulations. The Figure reveals a 
fair agreement of models with observational data and predictions from 
the hierarchical scaling model (HS; see, e.g., 
Bonometto et al. 1995, and references therein), 
which requires a constant $S_3$. In all cases a satisfactory fit is obtained 
with $S_3 \simeq 2.2$ (values are slightly higher for the
\L37 simulations than for C+2$\nu$DM,  but the difference is within
the errorbars).
For $R > 6\, h^{-1}$Mpc the values of $S_3$ decreases, rather
abruptly for \Mix\ and \La, gently in the other cases. The
significance of this trend is  questionable anyway in view of the large
uncertainties at these (relatively) large scales where the number of
sampling spheres is small and there may be effects due to the finite
size of the simulation box. Let us also recall that \Mix\ and \La\  have the
same initial random numbers.
Also $S_4$, though rather noisy, 
is compatible with HS by allowing a fit with a constant value in the 
range 6--7.
This rather good agreement of ``galaxies'' in the simulations with 
redshift survey results and HS confirms
and extends the results of Bonometto et al. (1995), and, in turn,
can be taken as indication that our galaxy identification procedure 
is a sensible one. 

It should be mentioned that 
the values of the reduced cumulants, $S_n=\bar \xi_n/\bar \xi_2^{n-1}$,
obtained from angular samples exceed those obtained from
redshift surveys by a factor of $\sim 3$ (Fry \& Gazta\~naga 1994,
Gazta\~naga 1994; see also Peebles, 1980).
The origin of this discrepancy is still unclear.
Since the galaxies included in angular catalogs 
span much larger volumes of
space than redshift surveys, it could be ascribed to 
sampling effects, i.e. that our local neighbourhood is not a fair sample 
(Gazta\~naga 1994), or finite statistics effects (Colombi, Bouchet \&
Schaeffer 1994 and 1995; Szapudi \& Colombi 1996). Indeed, the
last authors point out that the volumes of current redshift surveys and
the number of galaxies they contain appear to be too
small for a meaningful estimate of the $n$--point functions.
On the other hand, in our previous analysis of CDM and CHDM N--body 
simulations (Bonometto et al. 1995), we found
that the $S_n$ values are decreasing functions of the halo mass
cutoff for galaxy identification. Therefore, 
since projected samples include fainter galaxies, which could be
less biased 
tracers of the density field, this can partly account for the discrepancy 
between angular and redshift--space analyses. In any case,
if observational data and simulations 
are compared on strictly similar 
grounds, as we do, finite statistics should affect 
results for real and artificial galaxies by the same amount. However,
it is worth stressing here that the limits of our analysis 
should not be forgotten especially when comparing our results with
data from large angular samples.

Figures~5 and 6 give the results for the void probability function $P_0 (R)$
for simulated and PPS data. Errorbars for observational data 
are again 3$\sigma$ and the solid line represents
the expected behaviour for a Poisson sample with the same number of
objects as the real one. In Figure~5 the VPFs for different
$\delta_{th}$ are plotted.
In Figure~6, which also shows the effect of the sky variance,
the expected contribution to $P_0$ coming from Poisson noise is subtracted
and the resulting difference is divided by the volume $V(R) = (4\pi/3)R^3$.
Plotting $(P_0 - P_{0 ,Poisson})/V$ magnifies the detailed
behaviour of the VPF at small $R$. 

\ From Figures~5 and~6, it is clear that the C+2$\nu$DM model agrees  
with PPS data at all scales, independently of the choice of
$\delta_{th}$. In contrast, as before for $\bar\xi_2$, it is difficult for 
\L37 to yield an observer
setting whose sky is also marginally consistent with PPS data,
also when the overdensity threshold is pushed down to its lowest value 
$\delta_{th} = 100$. 

Figure 7 shows the results obtained for C$+2\nu$DM from artificial 
galaxy samples built starting from halo identification.
The 2--point function and VPF are shown for
the two galaxy identification methods described in the previous Section, 
applied to C+2$\nu$DM halos. Note that
the effect of halo fragmentation is rather limited, especially for
$P_0(R)$. This is essentially due to the presence of two effects, which
act in opposite directions in determining the strength of the "galaxy''
clustering. On the one hand, breaking up halos increases the mass
threshold. Therefore, galaxies are identified to correspond to higher
peaks of the DM density field, which in turn leads to an increase of their
clustering. On the other hand, since fragments generated by the same
halo are assigned different peculiar velocities, redshift--space
distortions cause a suppression of the clustering. The resulting stability of
$P_0(R)$ results can be also appreciated by comparing them with Figure 5. This
confirms that VPF results are connected with DM composition or model, while the
method of galaxy identification, within the class we considered here, which
is based on local and positive biasing, has only a modest relevance. 

In Figure~8 we report the behaviour of $P_1 (R)$ for data and models.
Observational
bootstrap errors are fairly wide here, especially at large $R$. 
In spite of that,
at $R < 4\, h^{-1}$Mpc, \L37 samples miss PPS data, while
C+2$\nu$DM is once more 
in good agreement with them. 
Similar considerations hold for the underdensity probability function
$U_\epsilon$,
which is illustrated in Figure~9 for a 30\% underdensity threshold.
Notice that, because of the point--like nature of the distribution,
$U_{\epsilon}$ carries new information with respect to $P_0$ only when 
$R$ approaches the average inter--particle separation, precisely when
$R \ge [300/(4\pi\epsilon)]^{1/3}\,d_{\rm gal}=3.42(\epsilon/100)^{-1/3}
\Mpch$. This is the reason why we plot results on $U_{30}$ only for
$R \magcir 4.5\,h^{-1}$Mpc.

Figure~10 is finally aimed to illustrate the effects of changing the
sampling of galaxies in overdensities (we take $\dth=150$ in all panels). 
Here we report the results
for $P_0$ and $P_1$, whose measure is potentially most sensitive
to the sampling choice. In each panel the dashed curve 
shows the scatter between different observer
settings in the ``usual'' realization, i.e. the one used to draw the
previous figures.
We also plot 1$\sigma$ errorbars that we obtain from such results.
Superimposed to them solid lines give the
results for two typical observer settings ``observing'' a different 
subset of particles, thus 
showing the limited effects of changing realization. In fact, 
the difference between the averages over 5 observers in two
different
realizations is smaller than the difference among observer settings in a
single realization by a factor $\sim 3$--10.

As a general remark, it can be said that {\sl cosmic variance} does not
appear to play an important role (an idea of its effect can be obtained 
by comparing \La\ and \Lb). In contrast, 
comparing  \Lc\ with its smaller--box
companions, there are non--neglegible differences. 
Since effects of finite box--size are expected to play a role on large
($\magcir 10\,h^{-1}$Mpc), differences on few Mpc scales are unlikely
to be directly related to the size of the simulation box.
For instance, Kauffmann \& Melott (1993) pointed out that the scaling of the
VPF starts feeling the box limits at about $L/4$. Therefore, any difference
on scales $\simeq 2$--$5\Mpch$, where we are mostly able to discriminate 
between models, seems more likely to be an effect of different resolutions:
in \Lc\ the linear cell size is a factor of $1.6$ larger and the mass of 
each DM particle is increased by a factor of $1.6^3\simeq4.1$.

\section{Conclusions}

In this work we tested the statistical properties of artificial galaxy samples
extracted from high--resolution simulations of C$+2\nu$DM with 
$\Omega_0=1$, $\Omega_\nu=0.2$, $h=0.5$ 
and $\Lambda$CDM with $\Omega_0=0.3$, $\Omega_\Lambda=0.7$, $h=0.7$ (\L37)
against a similar volume--limited sample of the 
PPS Survey. Artificial galaxies
reside in overdensity regions of the evolved density field whose density
contrasts are above a suitable threshold $\delta_{\rm th}$. 
We showed that, while the reduced skewness $S_3$  
yields almost identical results for the two models (and so does
kurtosis $S_4$ but with larger uncertainties), variance, $P_0$
and $P_1$  are able to discriminate
efficiently between them (also $U_\epsilon$, though marginally). 
In particular, the C$+2\nu$DM model 
agrees with our observational data, while it is
quite difficult to find an observer setting from which \L37 is
consistent with PPS data. The latter results confirm the 
analysis of Klypin, Primack \& Holtzman (1995) and
show that the excessive
small--scale clustering of \L37 is apparent in redshift--space
as well and makes this model hardly viable,
at least as long as galaxies follow DM overdensities.
In contrast, the analysis of $S_3$ (and $S_4$) does not
distinguish between the models, as said before, and agrees with PPS data
and HS predictions with constant values of $\simeq 2.2$ (and 6--7). 
This extends the results of Bonometto et al. (1995), which
also found a good agreement  of ``galaxies'' with HS in CDM and CHDM 
$N$--body simulations at variance with DM particles.

The values of $S_3$ and $S_4$ that we found agree with those 
derived from other redshift surveys, which,
as is known, are markedly smaller than those obtained from angular samples 
(see, e.g., Fry \& Gazta\~naga 1994).
As already stressed before, 
to address the origin of this discrepancy is beyond the
scope of this paper. However, we would like to notice here that
the remarkable stability of our results seems to indicate that sampling
effects do not play an important role. If redshift distortions,
projection effects and the mixing of galaxies of largely different
luminosities do not contribute either 
(see, e.g., Fry \& Gazta\~naga 1994 and Gazta\~naga 1994, who however used 
 the shallow CfA1 sample), the reason could very likely
be finite statistics effects, which indeed tend to decrease the estimates of
the hierarchical coefficients (Colombi, Bouchet \& Schaffer 1994; 
see also Szapudi \& Colombi 1996). This should not be a cause of concern 
for our analysis since we compare observational data and simulations
through ``galaxy'' samples of equal geometry, volume and inter--particle
separation. Even smaller effects are expected for the VPF,  
which has been found to be less sensitive to finite--volume effects
(Colombi, Bouchet \& Schaffer 1995).

Our analysis shows that \L37 tends to overproduce low--density regions. 
This is shown both by $P_0$ and $U_{30}$. Also, the probability of
finding a single galaxy in volumes smaller than $\sim 3$--$4\, h^{-1}$Mpc is
smaller than in the PPS data. 
These inconsistencies are more or less relevant in
various realizations, and depend on the threshold selected, but are present
everywhere. It seems clear that the galaxy number distribution in random
spheres is significantly different in \L37 and in the real world. 
However, let us add a word of caution 
about our conclusions in view of the limitations of our analysis, especially 
those related to the uncertainties 
on how galaxies actually form and on the way their real distribution
relates to that of DM particles.

As a concluding remark, it is worth pointing out what we have learned here 
about the ultimate goal of picking up the ``final'' cosmological model. 
As for the $\Lambda$CDM models, the one we considered here appears to have 
serious troubles in reproducing the galaxy clustering below $10\,h^{-1}$Mpc. It
is however clear that, by suitably changing the model 
parameters one may get substantial improvements (we reserve to a forthcoming 
paper the study of larger simulations of a larger suite of models).
As for the class of CHDM models, while 
the model with 1 massive neutrino providing $\Omega_\nu = 0.3$ 
fails to pass the VPF test (Ghigna et al. 1994), 
C+2$\nu$DM with $\Omega_\nu=0.2$ is in good agreement with all data
considered here. Therefore, having one single massive neutrino flavour 
with $m_\nu = 7\, $eV instead of two massive
neutrino flavours with $m_\nu = 2.3\, $eV seems completely
sufficient to alter the void distribution in a detectable way.
The remarkable performance of the C+2$\nu$DM model in this small--scale 
redshift--space analysis adds to
previous favorable results from numerical and linear theory calculations 
(see Primack et al. 1995, Primack 1996). This makes it a good
candidate to interpret the large scale structure of the Universe.

\acknowledgments   

AK and JRP acknowledge support from NASA
and NSF. AK and JRP utilized the CONVEX C3880 at the National Center for
Supercomputing Applications, University of Illinois at Urbana--Champaign.
SG would like to thank the University of California at Santa Cruz for
hospitality during the completion of this paper and Carlos~S. Frenk
for financial support.
The authors are grateful to Michael Vogeley for providing a computerized
version of his $\bar\xi_2$ and VPF results for the CfA2 catalog.

\clearpage

\begin{figure}
\caption{Comparison between results from CfA2 survey (open
circles; after Vogeley et al. 1994) and PPS both in LG (filled circles)
and CMB (triangles) frame. Left and right panels are for the variance
$\bar \xi_2(R)$ and the VPF $P_0(R)$, respectively. For reasons of
clarity, errorbars are reported only for PPS in LG frame and correspond
to $3\sigma$ bootstrap uncertainties.}
\end{figure}

\begin{figure}
\caption{Variance $\bar\xi_2$ {\sl vs.} scale $R$ for the set of
$800^3$--mesh simulations (continuous curves, which correspond to three 
 values of the overdensity threshold $\delta_{\rm th}$ as shown in the
first panel) and for
the PPS sample (filled circles; errorbars are $3\sigma$ bootstrap
errors). For each simulation and each $\delta_{\rm th}$, the curves are 
averages over 5 artificial samples differing in the observer location.
Of the three \L37 simulations, $\Lambda$CDM$_1$ has the same
initial random numbers as C$+2\nu$DM and a $50\Mpch$ box, 
$\Lambda$CDM$_2$ has the same box size but an independent set of random 
numbers, while $\Lambda$CDM${80}$ has a $80\Mpch$ box.}
\end{figure}

\begin{figure}
\caption{Effect of {\sl sky--variance} on $\bar\xi_2$. 
For each simulation, the curves correspond to 5  
different observer settings. PPS data are also shown as a reference.}
\end{figure}

\begin{figure}
\caption{Reduced Skewness $S_3\equiv \bar\xi_3/\bar\xi_2^2$ as 
a function of $R$ for simulated samples and PPS. Symbols are as in Figure 2.}
\end{figure}

\begin{figure}
\caption{Void probability function $P_0$ {\sl vs.} $R$ for
simulated samples and PPS. Symbols are the same as in Figure~2. In each panel, 
the solid curve is what is expected for a Poissonian
distribution of points with average separation $d_{\rm gal}$.}
\end{figure}

\begin{figure}
\caption{Here the VPF is plotted after  subtracting the poissonian 
$P_0$ and then dividing by the volume $V(R)$ of a sphere of radius $R$, 
which allows us to magnify the small scale behaviour. 
Each plot shows also $P_0 (R)$ for five typical different settings in the
simulations (dotted lines) and gives an indication of the sky variance. 
We have chosen the $\delta_{th}$s for which the $P_0$s of the models best 
approach the observational curve. In the top--left panel, the heavy ``T''
at the bottom sets the boundary of the region where the signal is
indistinguishable from Poissonian. They are obtained from the
3$\sigma$ scatters among measures for 50 different realizations of the 
Poissonian distribution in the same volume as our samples.}
\end{figure}

\begin{figure}
\caption{$\bar \xi_2(R)$ and $P_0(R)$ from
artificial samples based on halo identification. Dashed and dottes lines refer
to no fragmentation and full fragmentation of the DM halos, 
respectively, and are obtained as average over 10 realizations of artificial
samples. For sake of comparison, the result for PPS
(filled circles) is also given.}
\end{figure}

\begin{figure}
\caption{Results for $P_1(R)$, the probability of counting a
galaxy in a sphere of radius $R$. Symbols as in Figure~2.}
\end{figure}

\begin{figure}
\caption{Results for $U_{30}$, the probability that the number density
$n$ in a sphere of radius $R$ is less than 30\% of the average $\bar{n}$.}
\end{figure}

\begin{figure}
\caption{Effects of changing galaxy sampling in overdensities,
for $P_0$ and $P_1$. Dashed curves are results for different settings in the
realization used for the previous figures; 1$\sigma$ errorbars are worked out
from their variance. Continuous lines give results for two observer 
settings within a
different realization. Results for two models with $\delta_{\rm th}=150$ 
only are plotted.}
\end{figure}

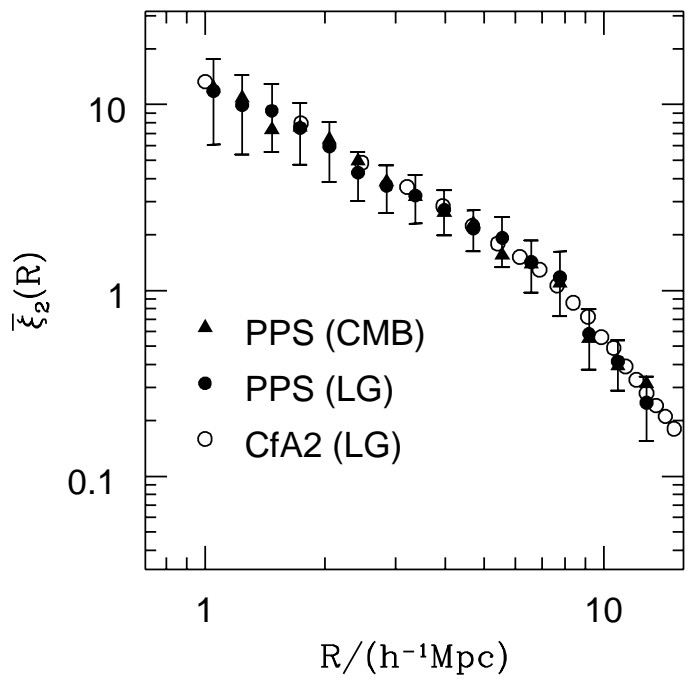
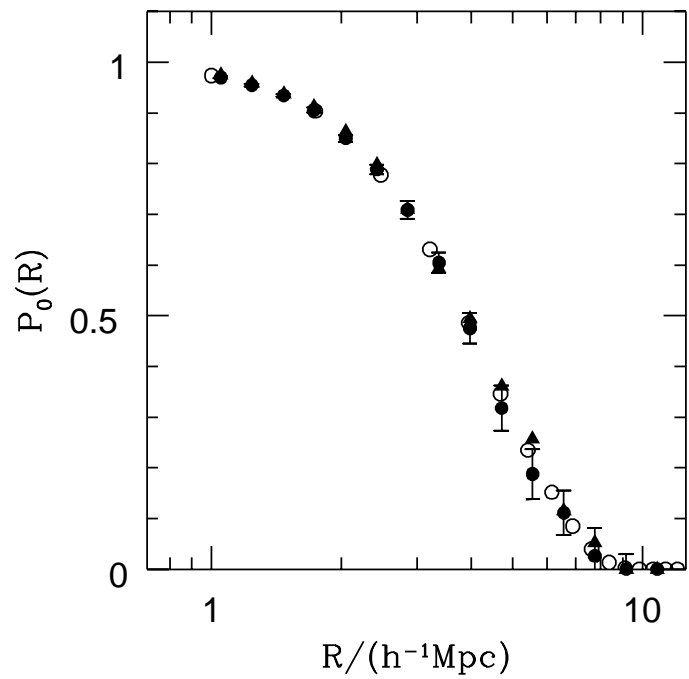
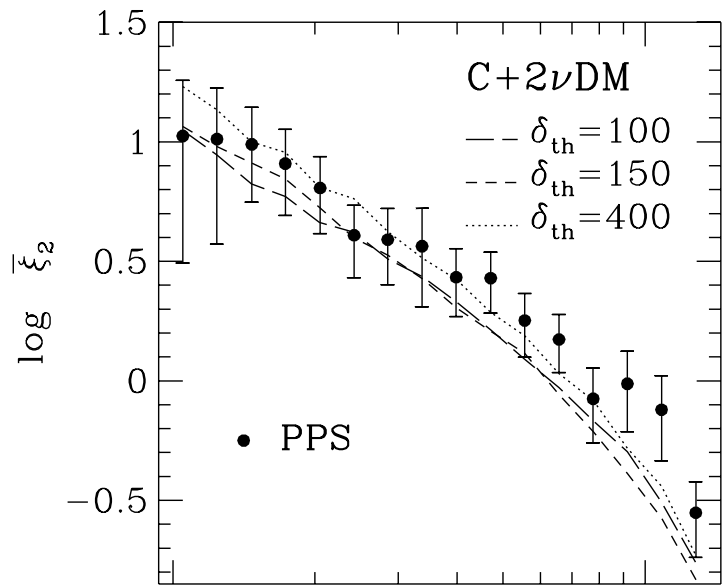
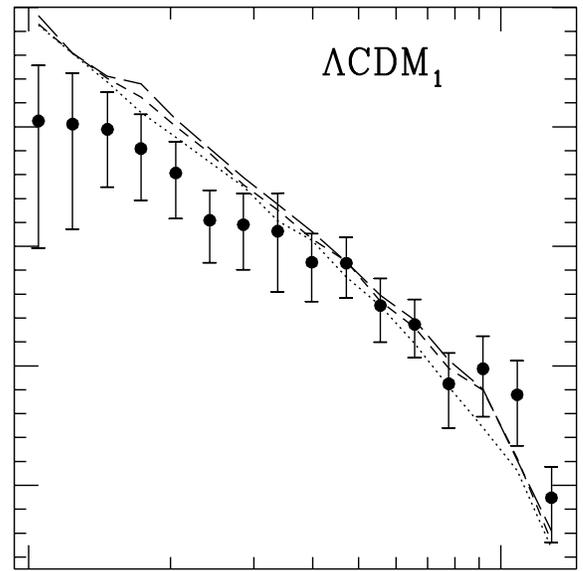
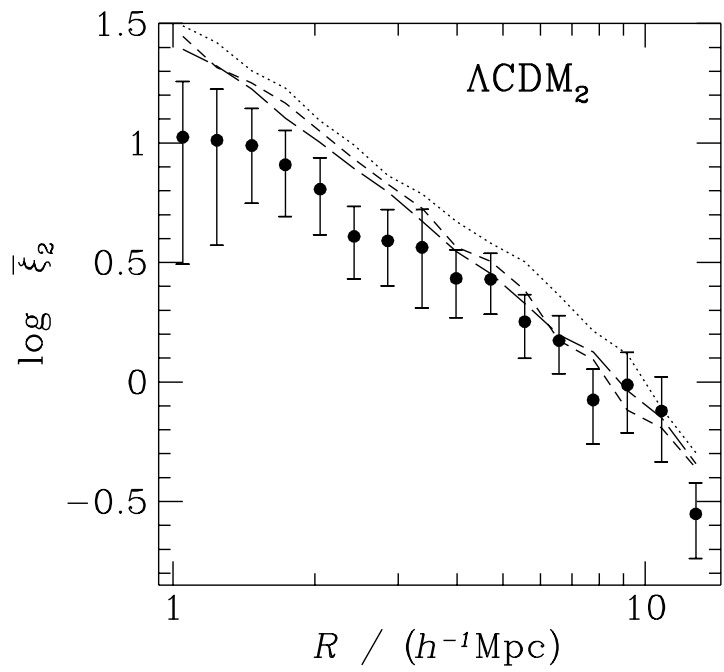
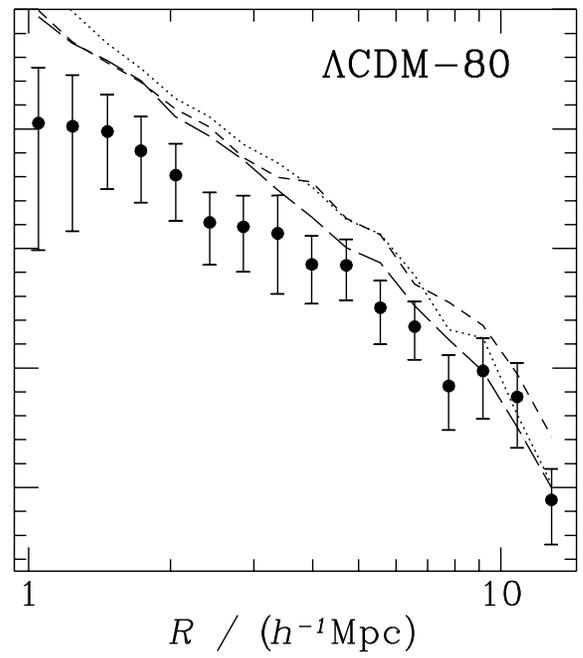
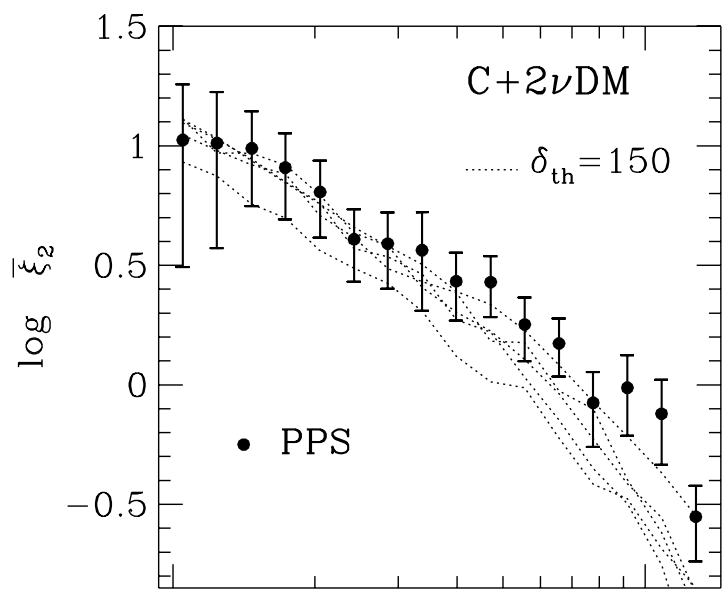
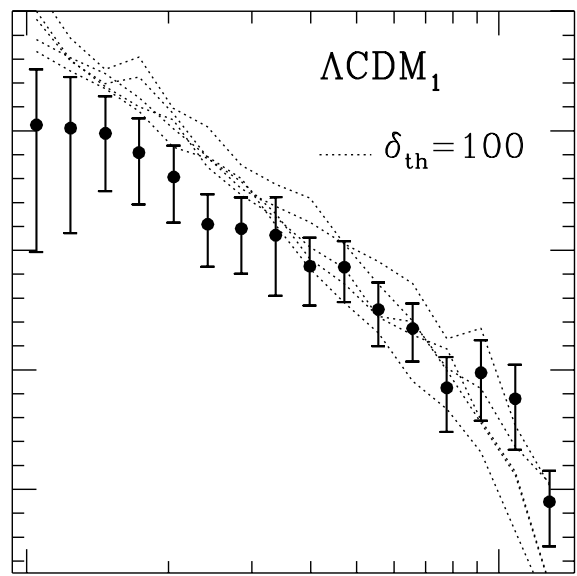
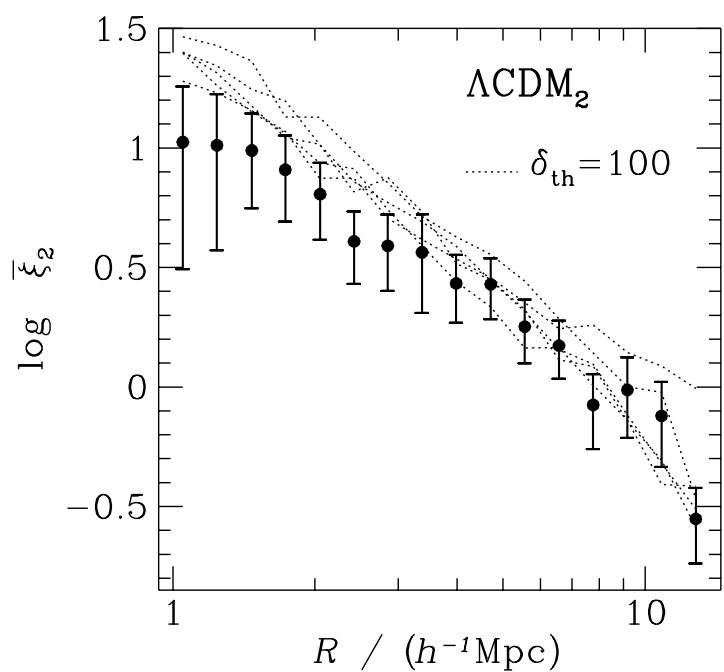
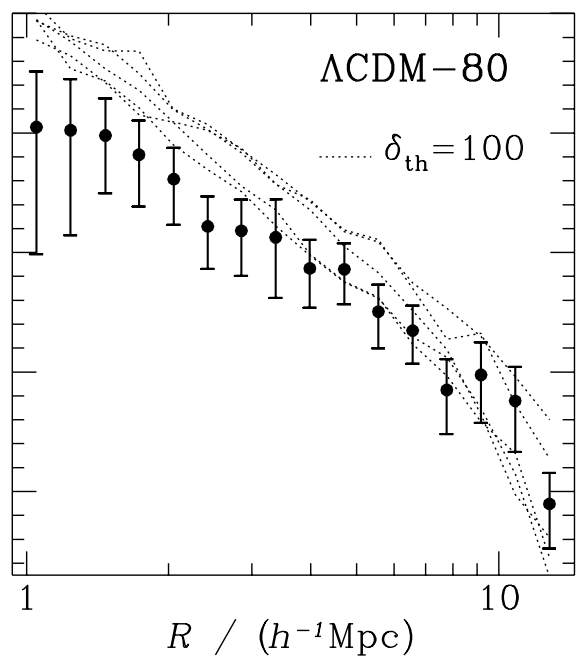
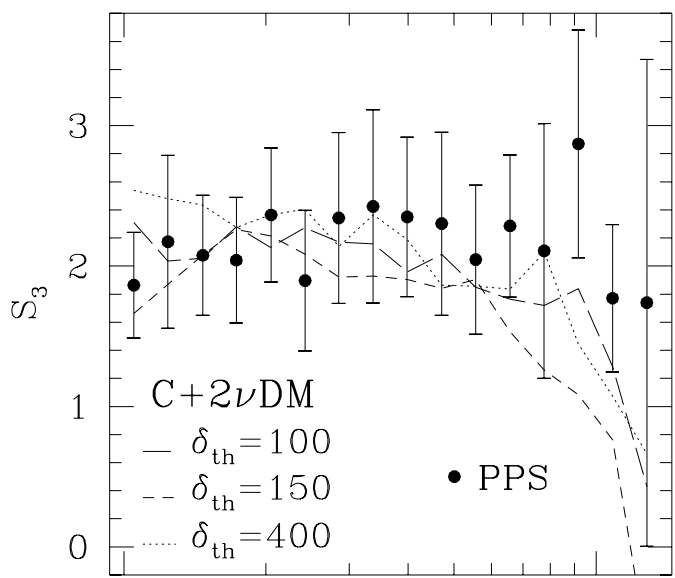
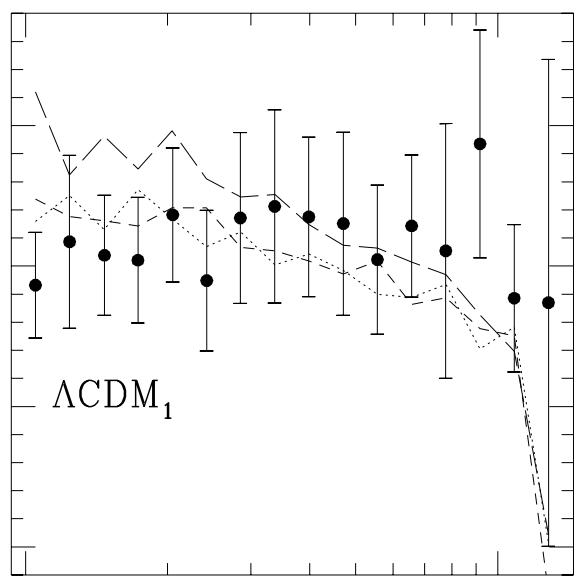
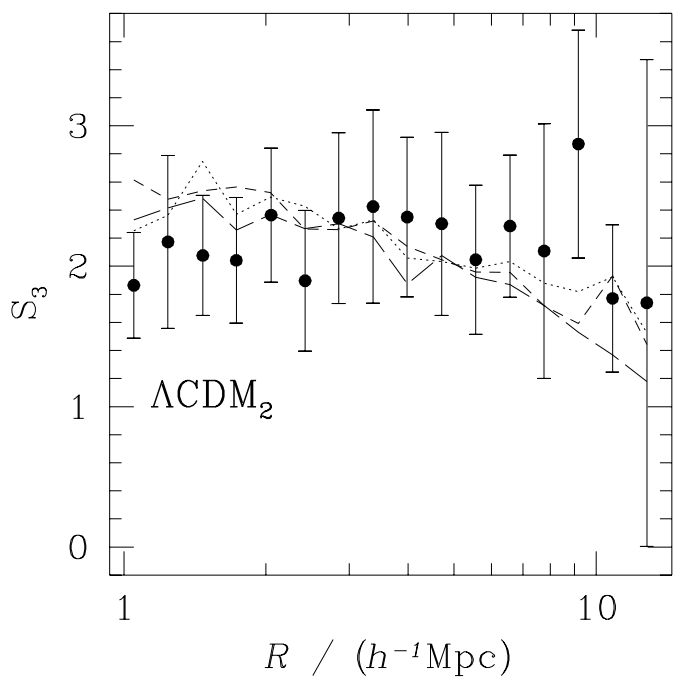
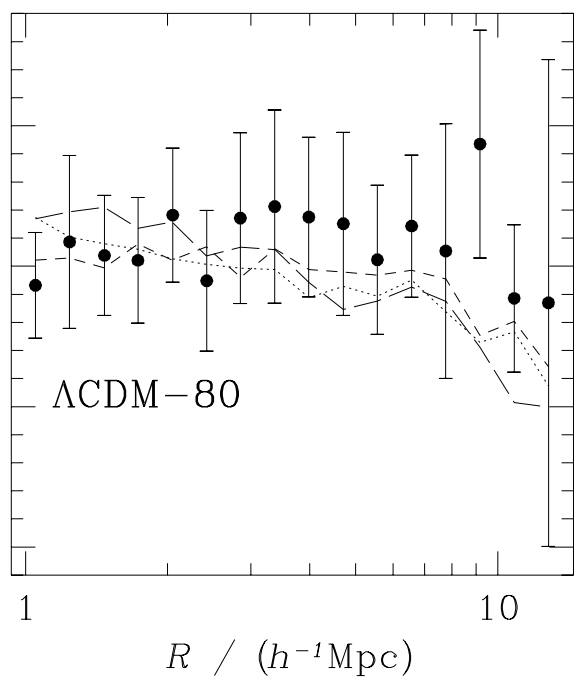
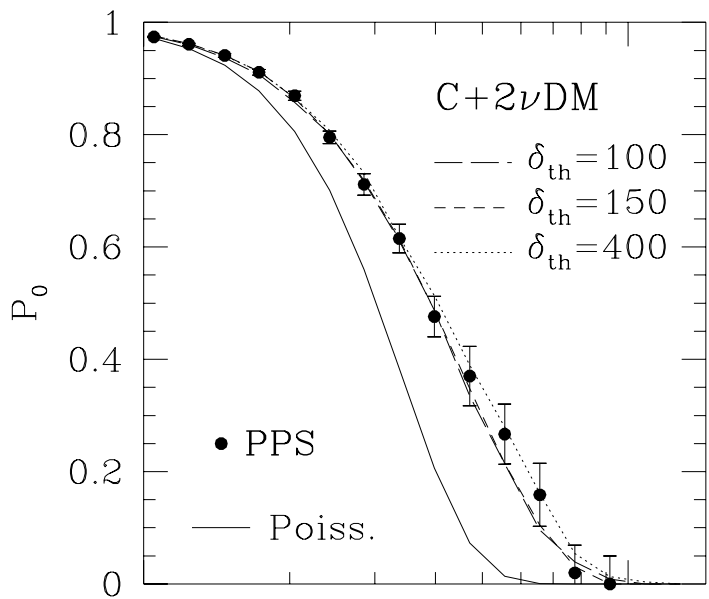
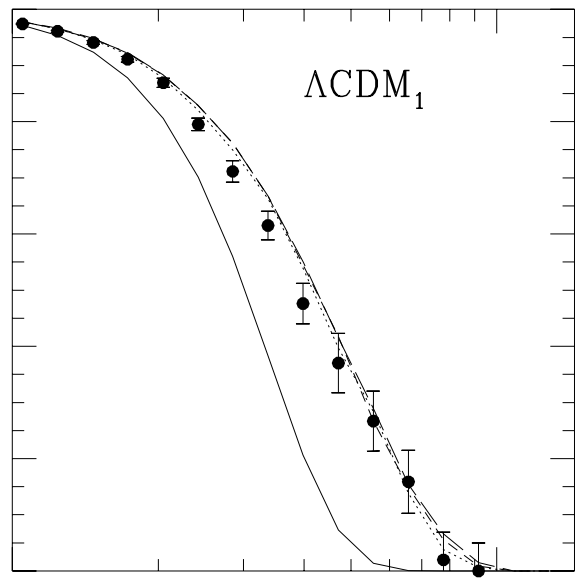
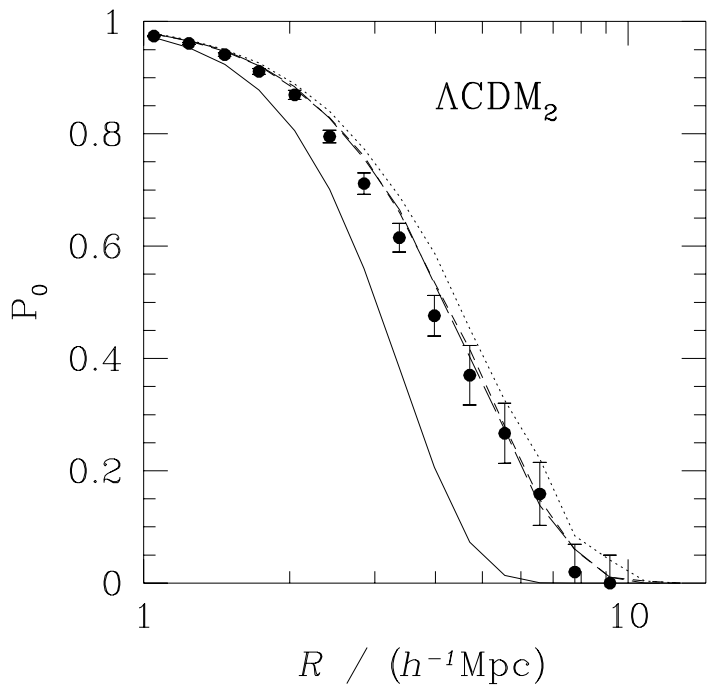
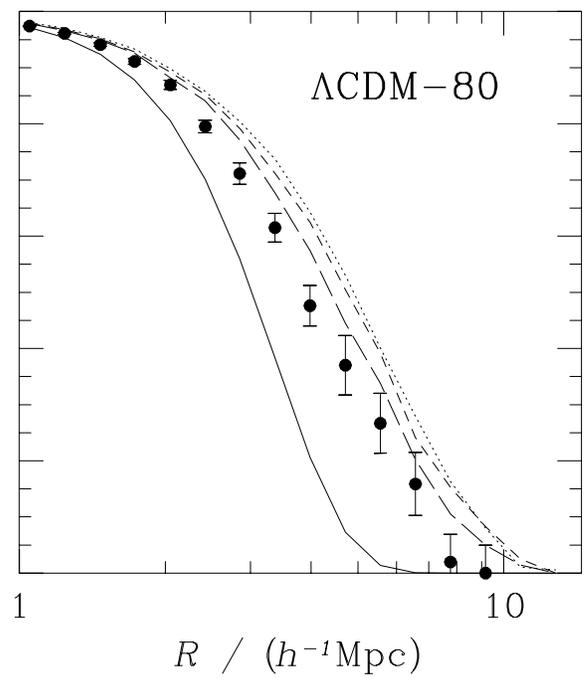
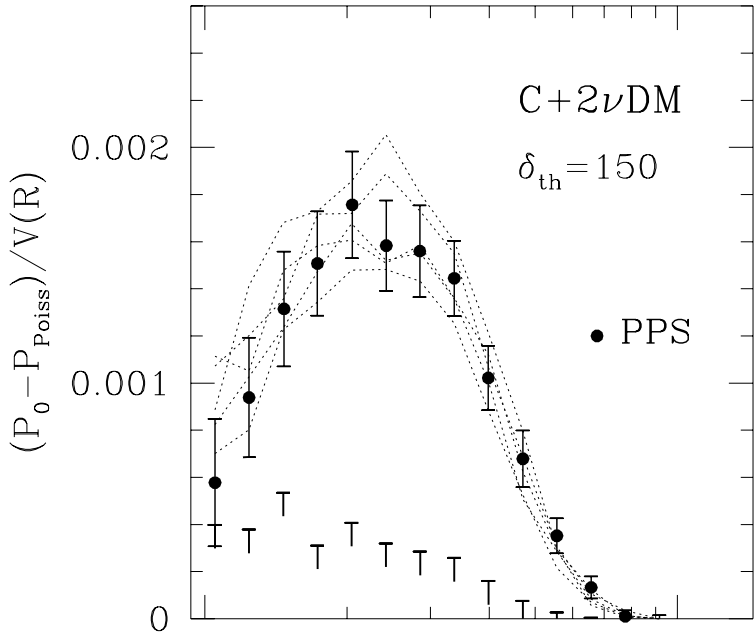
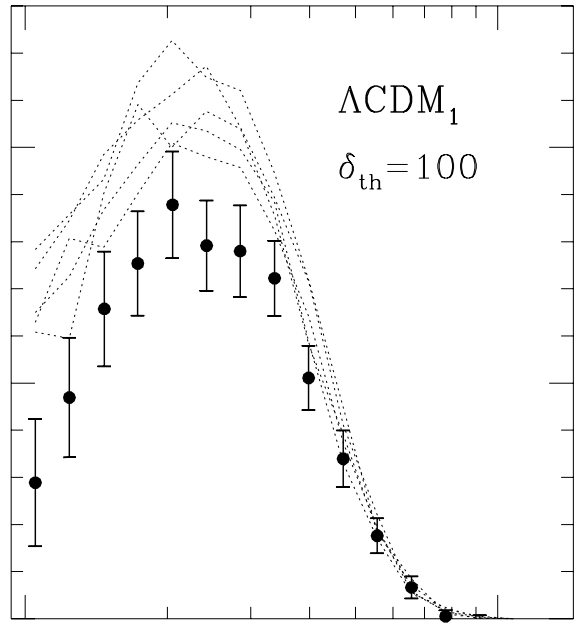
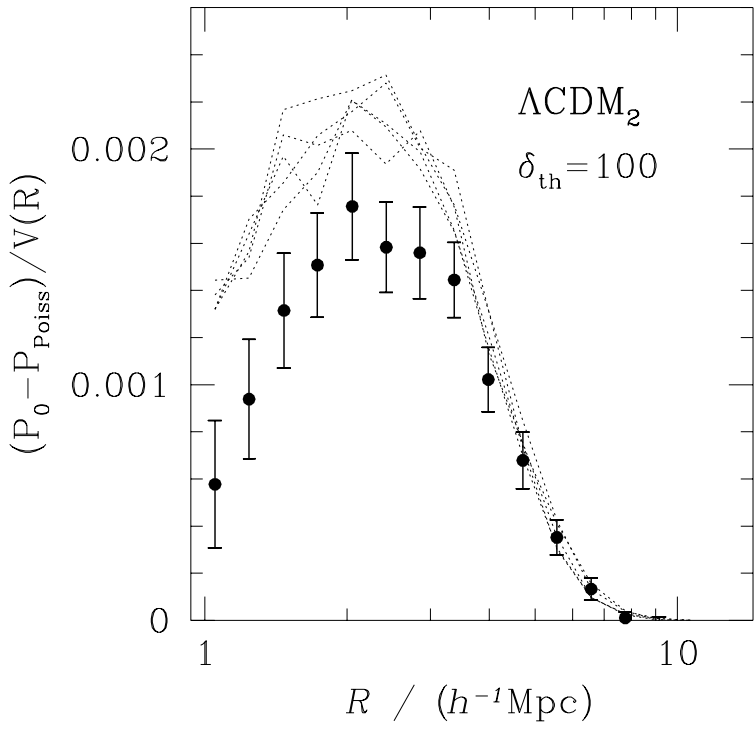
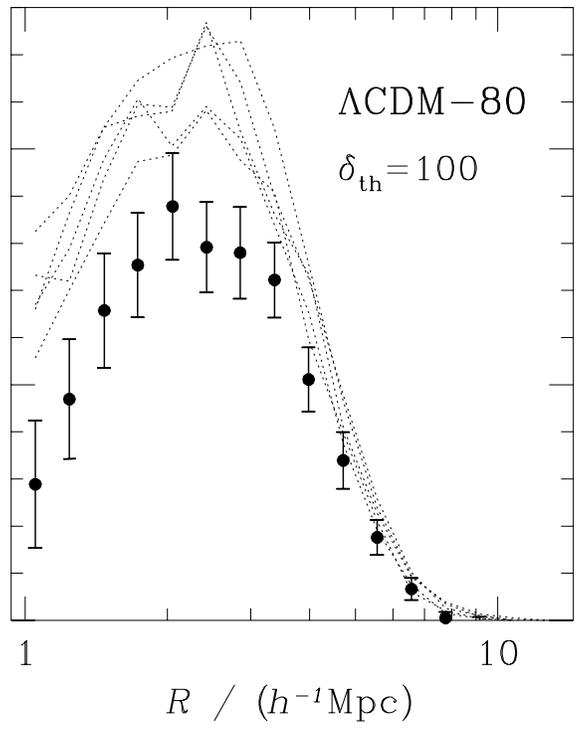
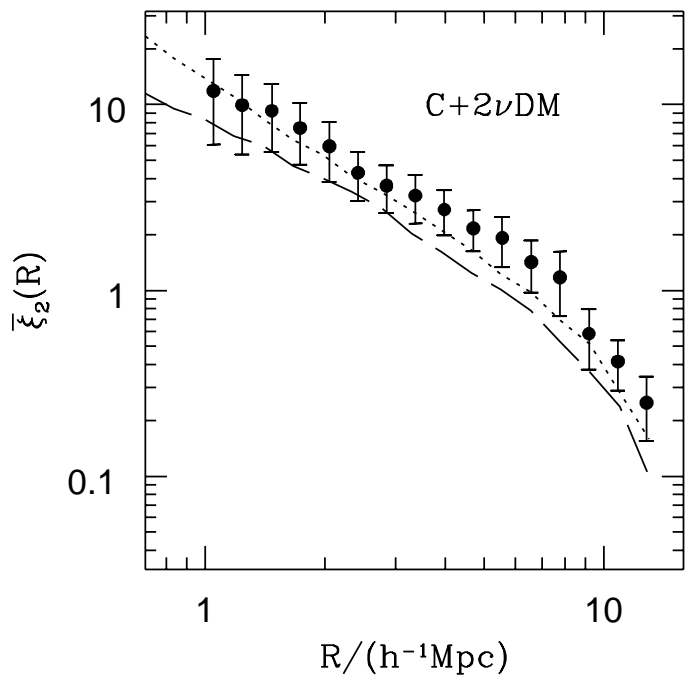
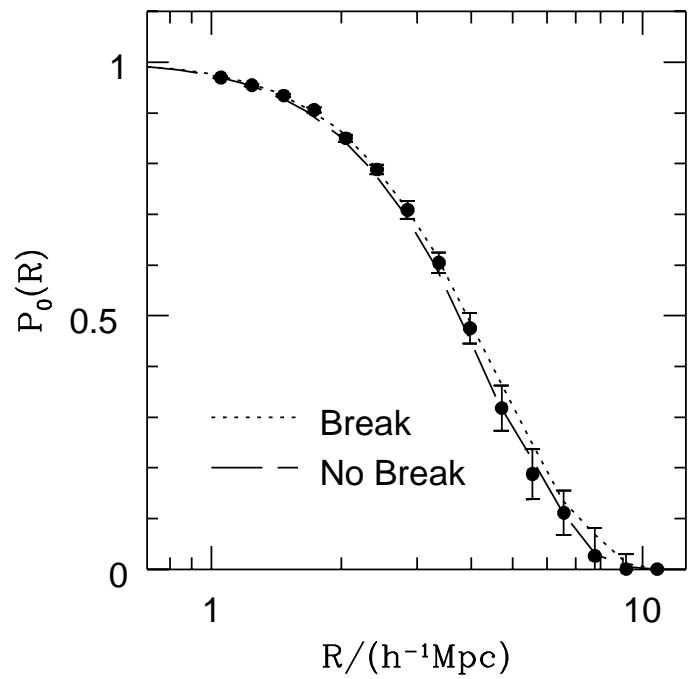
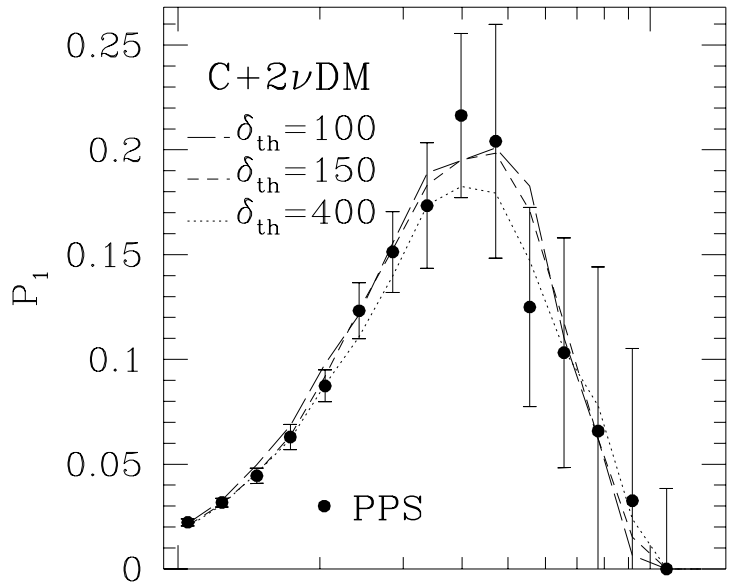
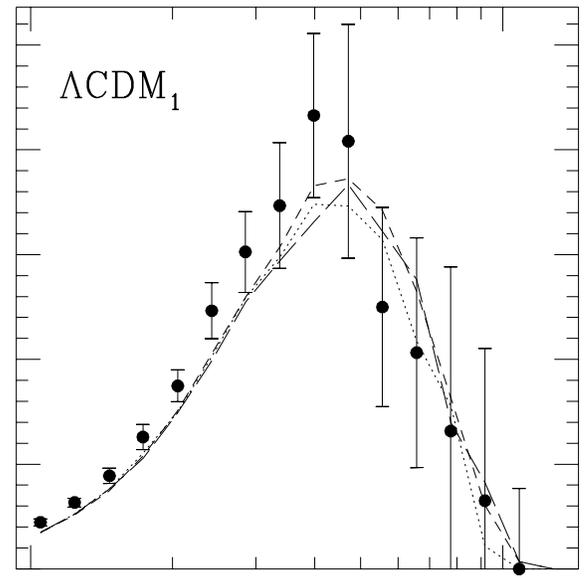
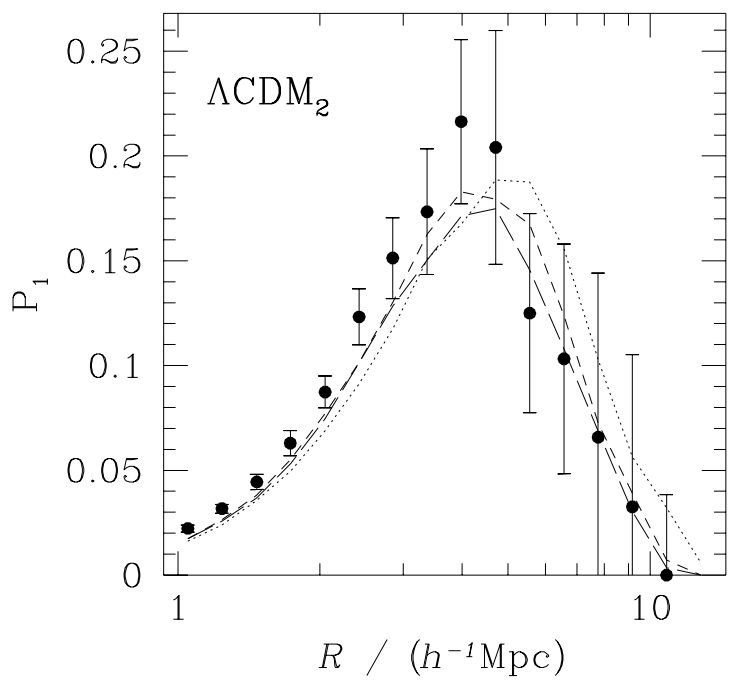
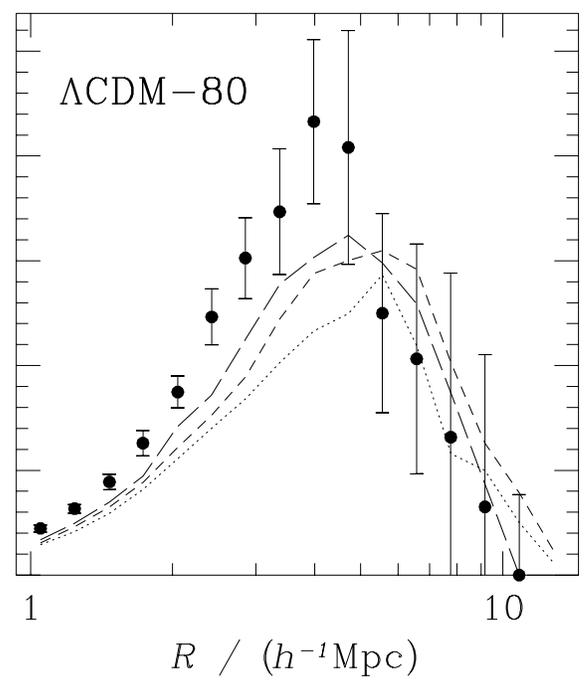
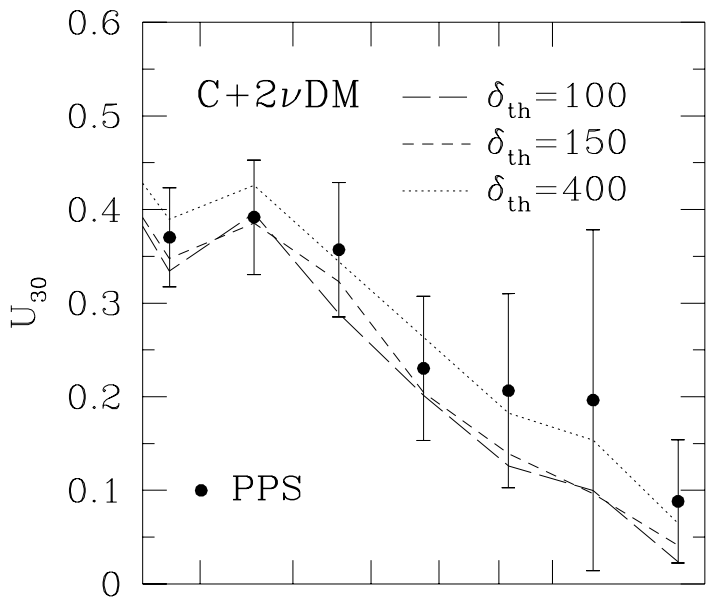
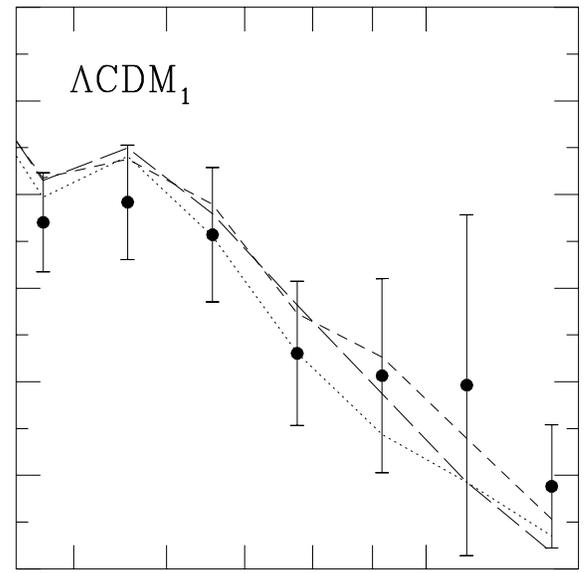
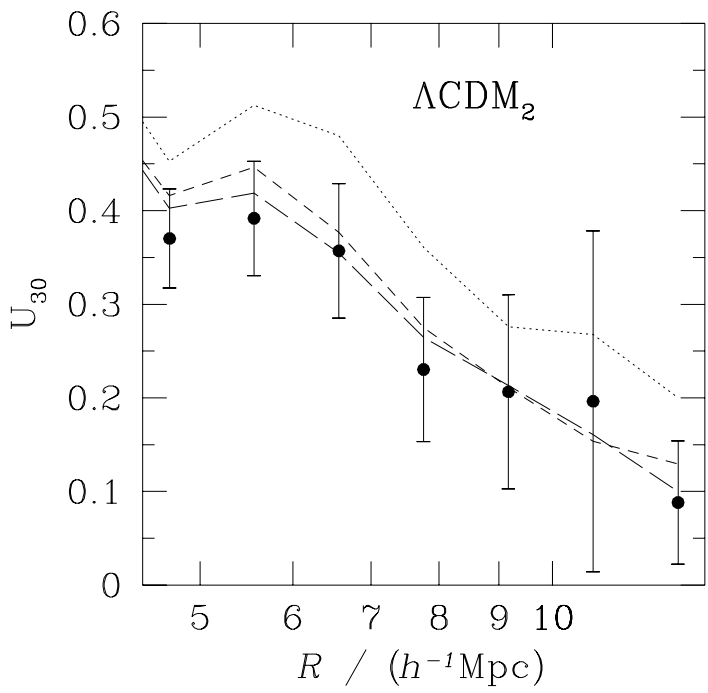
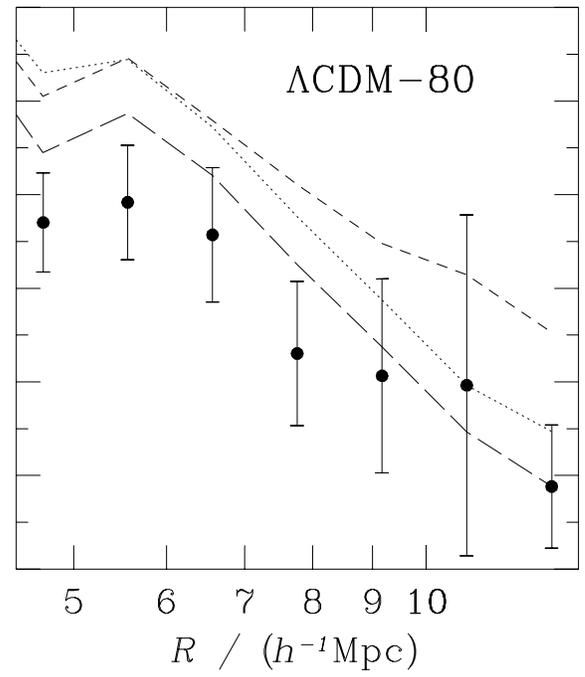
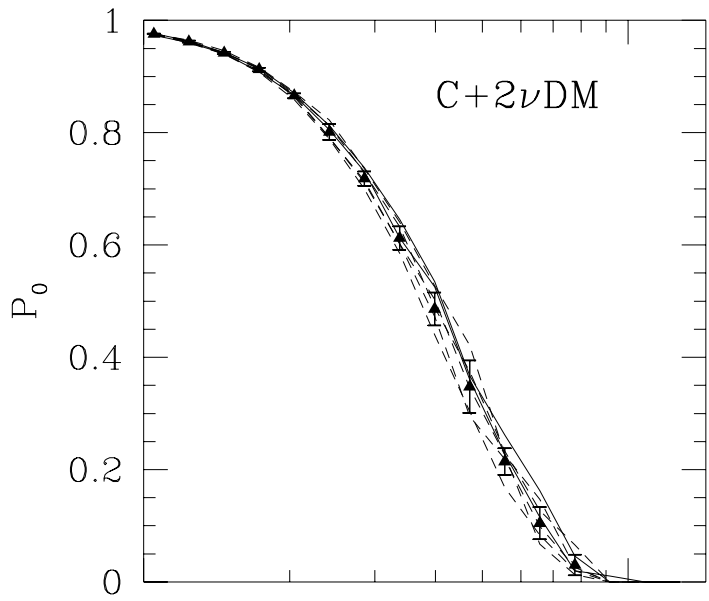
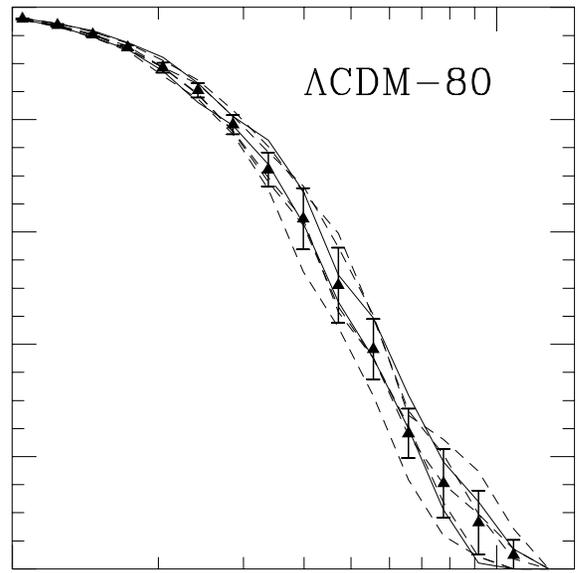
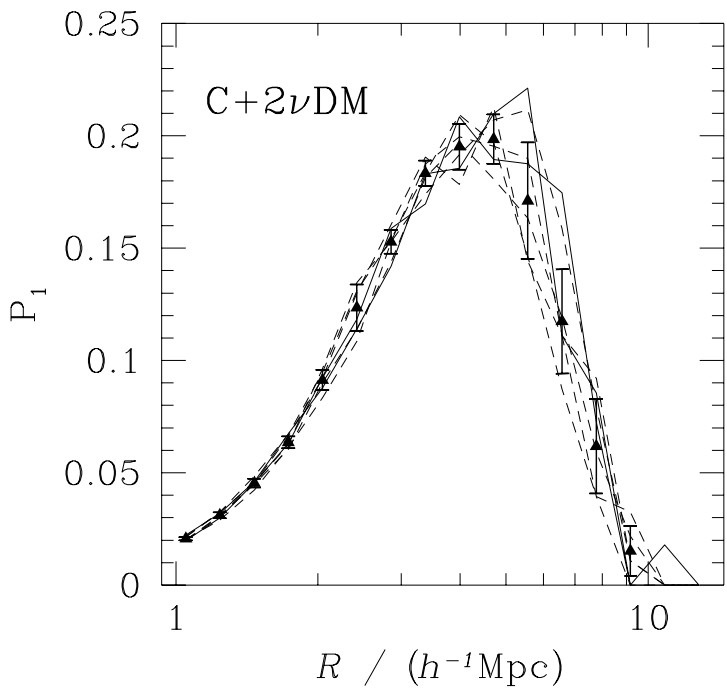
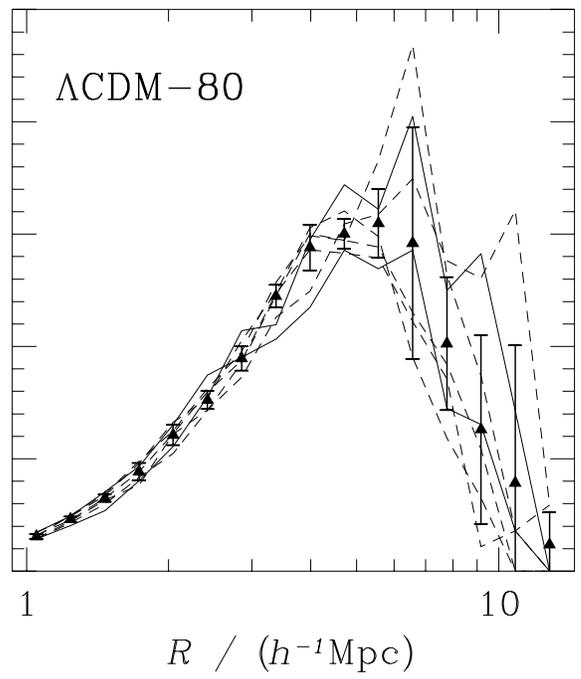
\end{document}